\documentclass[final,1p,times]{elsarticle}
\usepackage{amsfonts}
\usepackage{amsmath}
\usepackage{geometry}
\usepackage{bm}
\newtheorem{thm}{Theorem}

\newdefinition{rmk}{Remark}
\newproof{pf}{Proof}
\newproof{pot}{Proof of Theorem \ref{thm2}}
\usepackage{fancyhdr}
\begin{document}

\begin{frontmatter}
\title{Analytic solution for grand confluent hypergeometric function}

\author{Yoon Seok Choun
}
\ead{Yoon.Choun@baruch.cuny.edu; ychoun@gc.cuny.edu; ychoun@gmail.com}
\address{Baruch College, The City University of New York, Natural Science Department, A506, 17 Lexington Avenue, New York, NY 10010}
\begin{abstract}

In Ref.\cite{Chou2012a} I construct an approximative solution of the power series expansion in closed forms of Grand Confluent Hypergeometric (GCH) function only up to one term of $A_n$'s. And I obtain normalized constants and orthogonal relations of GCH function.

In this paper I will apply three term recurrence formula (3TRF) \cite{Chou2012b} to the power series expansion in closed forms of GCH function (for infinite series and polynomial which makes $B_n$ term terminated) including all higher terms of $A_n$'s. 

In general most of well-known special function with two recursive coefficients only has one eigenvalue for the polynomial case. However this new function with three recursive coefficients has infinite eigenvalues that make $B_n$'s term terminated at specific value of index $n$ because of 3TRF \cite{Chou2012b}. 

This paper is 9th out of 10 in series ``Special functions and three term recurrence formula (3TRF)''. See section 6 for all the papers in the series.  The previous paper in series deals with generating functions of Lame polynomial in the Weierstrass's form\cite{Chou2012h}. The next paper in the series describes the integral formalism and the generating function of GCH function\cite{Chou2012j}.
\end{abstract}

\begin{keyword}
Biconfluent Heun Equation, Three term recurrence formula, Asymptotic expansion 

\PACS{02.30.Hq, 02.30.Ik, 02.30.Gp, 03.65.Ge, 03.65.-w}
\end{keyword}
                                      
\end{frontmatter}  
\section{Introduction}
Biconfluent Heun (BCH) function, a confluent form of Heun function\cite{Heun1889,Ronv1995}, is the special case of Grand Confluent Hypergeometric (GCH) function\cite{Chou2012a}\footnote{For the canonical form of BCH equation \cite{Ronv1995}, replace $\mu $, $\varepsilon $, $\nu $, $\Omega $ and $\omega $ by $-2$, $-\beta  $, $ 1+\alpha $, $\gamma -\alpha -2 $ and $ 1/2 (\delta /\beta +1+\alpha )$ in (\ref{eq:1}). For DLFM version \cite{NIST} or in ref.\cite{Slavy2000}, replace $\mu $ and $\omega $ by 1 and $-q/\varepsilon $ in (\ref{eq:1}).}: this has a regular singularity at $x=0$, and an irregular singularity at $\infty$ of rank 2. For example, the BCH function is included in the radial Schr$\ddot{\mbox{o}}$dinger equation with rotating harmonic oscillator and a class of confinement potentials: recently it's started to appear in theoretical modern physics \cite{Slav1996,Ralk2002,Kand2005,Hortacsu:2011rr,Arri1991}. 

In Ref.\cite{Chou2012c,Chou2012d}, I construct power series expansions in closed form and integral representations of Heun equation by applying 3TRF. Heun equation is applicable to diverse areas such as theory of black holes, lattice systems in statistical mechanics, addition of three quantum spins, solutions of the Schr$\ddot{\mbox{o}}$dinger equation of quantum mechanics. \cite{Hortacsu:2011rr,Take2008,Suzu1999,Suzu1998}

In Ref.\cite{Chou2012a}, I show an analytic solution of GCH equation only up to one term of $A_n$'s. In this paper I construct the power series expansion of GCH equation in closed forms and asymptotic behaviors including all higher terms of $A_n$'s by applying 3TRF \cite{Chou2012b}. 
\begin{equation}
x \frac{d^2{y}}{d{x}^2} + \left( \mu x^2 + \varepsilon x + \nu  \right) \frac{d{y}}{d{x}} + \left( \Omega x + \varepsilon \omega \right) y = 0
\label{eq:1}
\end{equation}
(\ref{eq:1}) is a Grand Confluent Hypergeometric (GCH) differential equation where $\mu$, $\varepsilon$, $\nu $, $\Omega$ and $\omega$ are real or imaginary parameters.\cite{Chou2012a} It has a regular singularity at the origin and an irregular singularity at the in¯infinity. Biconfluent Heun Equation is derived, the special case of GCH equation, by putting coefficients $\mu =1$ and $\omega =-q/\varepsilon $.\cite{NIST} 

$y(x)$ has a series expansion of the form
\begin{equation}
y(x)= \sum_{n=0}^{\infty } c_n x^{n+\lambda }
\label{eq:2}
\end{equation}
where $\lambda$ is an indicial root. Plug (\ref{eq:2}) into (\ref{eq:1}).
\begin{equation}
c_{n+1}=A_n \;c_n +B_n \;c_{n-1} \hspace{1cm};n\geq 1
\label{eq:3}
\end{equation}
where,
\begin{subequations}
\begin{equation}
A_n = -\frac{\varepsilon (n+\omega +\lambda )}{(n+1+\lambda )(n+\nu +\lambda )}
\label{eq:4a}
\end{equation}
\begin{equation}
B_n = -\frac{\Omega +\mu (n-1+\lambda )}{(n+1+\lambda )(n+\nu +\lambda )}
\label{eq:4b}
\end{equation}
\begin{equation}
c_1= A_0 \;c_0
\label{eq:4c}
\end{equation}
\end{subequations}
We have two indicial roots which are $\lambda = 0$ and $ 1-\nu $
\section{\label{sec:level2}Power series}

\subsection{A polynomial which makes $B_n$ term terminated}
There are three types of polynomials in three term recurrence relation of a linear ordinary differential equation: (1) a polynomial which makes $B_n$ term terminated: $A_n$ term is not terminated, (2) a polynomial which makes $A_n$ term terminated: $B_n$ term is not terminated, (3) a polynomial which makes $A_n$ and $B_n$ terms terminated at the same time.\footnote{If $A_n$ and $B_n$ terms are not terminated, it turns to be infinite series.} In general the GCH (or Biconfluent Heun) polynomial is defined as type 3 polynomial where $A_n$ and $B_n$ terms terminated. The GCH polynomial comes from GCH equation that has fixed values of $\Omega $ and $\omega $. In three term recurrence relation, a polynomial of type 3 I categorize as a complete polynomial. In this paper I construct power series expansions and asymptotic series in closed forms for the GCH polynomial which makes $B_n$ term terminated: I treat $\varepsilon $, $\mu $, $\nu $ and $\omega $ as free variables and a parameter $\Omega$ as a fixed value. In chapter 6 of Ref.\cite{Choun2013}, I construct formal series solutions in closed forms and integrals of the GCH polynomial which makes $A_n$ term terminated including generating functions of it analytically: I treat $\varepsilon $, $\mu $, $\nu $ and $\Omega $ as free variables and a parameter $\omega$ as a fixed value. In future papers I will derive a type 3 GCH polynomial.
\begin{thm}
In Ref.\cite{Chou2012b}, the general expression of a power series of $y(x)$ for a polynomial which makes $B_n$ term terminated is defined by
\begin{eqnarray}
 y(x)&=& \sum_{n=0}^{\infty } y_{n}(x)=  y_0(x)+ y_1(x)+ y_2(x)+y_3(x)+\cdots \nonumber\\
&=& c_0 \Bigg\{ \sum_{i_0=0}^{\beta _0} \left( \prod _{i_1=0}^{i_0-1}B_{2i_1+1} \right) x^{2i_0+\lambda } \nonumber\\
&&+ \sum_{i_0=0}^{\beta _0}\left\{ A_{2i_0} \prod _{i_1=0}^{i_0-1}B_{2i_1+1}  \sum_{i_2=i_0}^{\beta _1} \left( \prod _{i_3=i_0}^{i_2-1}B_{2i_3+2} \right)\right\} x^{2i_2+1+\lambda }\nonumber\\
 && + \sum_{N=2}^{\infty } \Bigg\{ \sum_{i_0=0}^{\beta _0} \Bigg\{A_{2i_0}\prod _{i_1=0}^{i_0-1} B_{2i_1+1} \prod _{k=1}^{N-1} \Bigg( \sum_{i_{2k}= i_{2(k-1)}}^{\beta _k} A_{2i_{2k}+k}\prod _{i_{2k+1}=i_{2(k-1)}}^{i_{2k}-1}B_{2i_{2k+1}+(k+1)}\Bigg)\nonumber\\
 &&\times  \sum_{i_{2N} = i_{2(N-1)}}^{\beta _N} \Bigg( \prod _{i_{2N+1}=i_{2(N-1)}}^{i_{2N}-1} B_{2i_{2N+1}+(N+1)} \Bigg) \Bigg\} \Bigg\} x^{2i_{2N}+N+\lambda }\Bigg\}
  \label{eq:5}
\end{eqnarray}
For a polynomial, we need a condition:
\begin{equation}
 B_{2\beta _i + (i+1)}=0 \hspace{1cm} \mathrm{where}\; i,\beta _i \in \mathbb{N}_{0}
 \label{eq:6}
\end{equation}
\end{thm}
In this paper Pochhammer symbol $(x)_n$ is used to represent the rising factorial: $(x)_n = \frac{\Gamma (x+n)}{\Gamma (x)}$.
On the above $ \beta _i$ is an eigenvalue that makes $B_n$ term terminated at certain value of index $n$. (\ref{eq:6}) makes each $y_i(x)$ where $i \in \mathbb{N}_{0}$ as the polynomial in (\ref{eq:5}). Substitute (\ref{eq:4a})-(\ref{eq:4c}) into (\ref{eq:5}) by using (\ref{eq:6}). 
The general expression of a power series of GCH equation for a polynomial which makes $B_n$ term terminated is given by
\begin{eqnarray}
 y(x)&=& \sum_{n=0}^{\infty } y_{n}(x)=  y_0(x)+ y_1(x)+ y_2(x)+y_3(x)+\cdots \nonumber\\
&=&  c_0 x^{\lambda } \Bigg\{\sum_{i_0=0}^{\beta _0} \frac{(-\beta _0)_{i_0}}{(1+\frac{\lambda }{2})_{i_0}(\gamma +\frac{\lambda }{2})_{i_0}}z^{i_0} \nonumber\\
&&+   \Bigg\{\sum_{i_0=0}^{\beta _0} \frac{(i_0+\frac{\lambda }{2}+\frac{\omega }{2})}{(i_0+\frac{1}{2}+\frac{\lambda }{2})(i_0-\frac{1}{2}+\gamma +\frac{\lambda }{2})} \frac{(-\beta _0)_{i_0}}{(1+\frac{\lambda }{2})_{i_0}(\gamma +\frac{\lambda }{2})_{i_0}}\nonumber\\
&&\times  \sum_{i_1=i_0}^{\beta _1} \frac{(-\beta _1)_{i_1}(\frac{3}{2}+\frac{\lambda }{2})_{i_0}(\gamma +\frac{1}{2}+ \frac{\lambda }{2})_{i_0}}{(-\beta _1)_{i_0}(\frac{3}{2}+\frac{\lambda }{2})_{i_1}(\gamma +\frac{1}{2}+\frac{\lambda }{2})_{i_1}} z^{i_1} \Bigg\} \tilde{\varepsilon }\nonumber
\end{eqnarray}
\begin{eqnarray}
&&+ \sum_{n=2}^{\infty } \Bigg\{ \sum_{i_0=0}^{\beta _0}  \frac{(i_0+\frac{\lambda }{2}+\frac{\omega }{2})}{(i_0+\frac{1}{2}+\frac{\lambda }{2})(i_0-\frac{1}{2}+\gamma +\frac{\lambda }{2})} \frac{(-\beta _0)_{i_0}}{(1+\frac{\lambda }{2})_{i_0}(\gamma +\frac{\lambda }{2})_{i_0}} \nonumber\\
&&\times \prod _{k=1}^{n-1} \Bigg\{ \sum_{i_k=i_{k-1}}^{\beta _k} \frac{(i_k+\frac{\lambda }{2}+\frac{\omega }{2}+\frac{k}{2})}{(i_k+\frac{1}{2}+\frac{\lambda }{2}+\frac{k}{2})(i_k-\frac{1}{2}+\gamma + \frac{k}{2}+\frac{\lambda }{2})} \nonumber\\
&&\times \frac{(-\beta _k)_{i_k}(1+\frac{k}{2}+\frac{\lambda }{2})_{i_{k-1}}(\frac{k}{2}+\gamma +\frac{\lambda }{2})_{i_{k-1}}}{(-\beta _k)_{i_{k-1}}(1+\frac{k}{2}+\frac{\lambda }{2})_{i_k}(\frac{k}{2}+\gamma +\frac{\lambda }{2})_{i_k}}\Bigg\} \nonumber\\
&&\times \sum_{i_n= i_{n-1}}^{\beta _n} \frac{(-\beta _n)_{i_n}(1+\frac{n}{2}+\frac{\lambda }{2})_{i_{n-1}}(\frac{n}{2}+\gamma +\frac{\lambda }{2})_{i_{n-1}}}{(-\beta _n)_{i_{n-1}}(1+\frac{n}{2}+\frac{\lambda }{2})_{i_n}(\frac{n}{2}+\gamma +\frac{\lambda }{2})_{i_n}} z^{i_n}\Bigg\} \tilde{\varepsilon }^n \Bigg\}
\label{eq:7}
\end{eqnarray}
where
\begin{equation}
\begin{cases} z = -\frac{1}{2}\mu x^2 \cr
\tilde{\varepsilon }  = -\frac{1}{2}\varepsilon  x\cr
\gamma  = \frac{1}{2}(1+\nu ) \cr
\Omega = -\mu (2\beta _i+i+\lambda )\;\;\mbox{as}\;i,\beta_i  \in \mathbb{N}_{0}\cr
\mbox{As}\; \beta _i\leq \beta _j\;\;\mbox{only}\;\;\mbox{if}\;\;i\leq j
\end{cases}
\label{eq:8}
\end{equation}
Put $c_0$= $\frac{\Gamma (\gamma +\beta _0)}{\Gamma (\gamma )}$ as $\lambda $=0 in (\ref{eq:7}).
\begin{rmk}
The power series expansion of GCH equation of the first kind for a polynomial which makes $B_n$ term terminated around $x=0 $ as $\Omega = -2\mu (\beta _i+\frac{i}{2})$ where $i, \beta _i \in \mathbb{N}_{0}$ is given by
\begin{eqnarray}
 y(x)&=& QW_{\beta _i}\left( \beta _i=-\frac{\Omega }{2\mu }-\frac{i}{2} , \omega,  \gamma =\frac{1}{2}(1+\nu );\; \tilde{\varepsilon }= -\frac{1}{2}\varepsilon x;\; z=-\frac{1}{2}\mu x^2 \right) \nonumber\\
&&= \frac{\Gamma (\gamma +\beta _0)}{\Gamma (\gamma )} \Bigg\{\sum_{i_0=0}^{\beta _0 } \frac{(-\beta _0)_{i_0}}{(1)_{i_0}(\gamma)_{i_0}}z^{i_0}+   \Bigg\{ \sum_{i_0=0}^{\beta _0 }\frac{(i_0+\frac{\omega }{2})}{(i_0+\frac{1}{2})(i_0-\frac{1}{2}+\gamma )} \frac{(-\beta  _0)_{i_0}}{(1)_{i_0}(\gamma)_{i_0}}\nonumber\\
&&\times  \sum_{i_1=i_0}^{\beta _1} \frac{(-\beta _1)_{i_1}(\frac{3}{2})_{i_0}(\gamma +\frac{1}{2})_{i_0}}{(-\beta _1)_{i_0}(\frac{3}{2})_{i_1}(\gamma +\frac{1}{2})_{i_1}} z^{i_1} \Bigg\} \tilde{\varepsilon }
+ \sum_{n=2}^{\infty } \Bigg\{ \sum_{i_0=0}^{\beta _0} \frac{(i_0+\frac{\omega }{2})}{(i_0+\frac{1}{2})(i_0-\frac{1}{2}+\gamma)} \frac{(-\beta _0)_{i_0}}{(1)_{i_0}(\gamma )_{i_0}} \nonumber\\
&&\times \prod _{k=1}^{n-1} \Bigg\{ \sum_{i_k=i_{k-1}}^{\beta _k} \frac{(i_k+\frac{\omega }{2}+\frac{k}{2})}{(i_k+\frac{1}{2}+\frac{k}{2})(i_k-\frac{1}{2}+\gamma + \frac{k}{2})} \frac{(-\beta _k)_{i_k}(1+\frac{k}{2})_{i_{k-1}}(\frac{k}{2}+\gamma )_{i_{k-1}}}{(-\beta _k)_{i_{k-1}}(1+\frac{k}{2})_{i_k}(\frac{k}{2}+\gamma )_{i_k}}\Bigg\} \nonumber\\
&&\times \sum_{i_n= i_{n-1}}^{\beta _n} \frac{(-\beta _n)_{i_n}(1+\frac{n}{2})_{i_{n-1}}(\frac{n}{2}+\gamma )_{i_{n-1}}}{(-\beta _n)_{i_{n-1}}(1+\frac{n}{2})_{i_n}(\frac{n}{2}+\gamma )_{i_n}} z^{i_n}\Bigg\} \tilde{\varepsilon }^n \Bigg\}\nonumber
\end{eqnarray}
\end{rmk}
For the minimum value of the GCH equation of the first kind for a polynomial which makes $B_n$ term terminated around $x=0 $, put $\beta _0=\beta _1=\beta _2=\cdots=0$ in Remark 1.
\begin{eqnarray}
 y(x)&=& QW_{0}\left( \Omega =-i\mu , \omega,  \gamma =\frac{1}{2}(1+\nu );\; \tilde{\varepsilon }= -\frac{1}{2}\varepsilon x;\; z=-\frac{1}{2}\mu x^2 \right) \nonumber\\
&=& \; _1F_1\left( \omega ,\nu ,-\varepsilon x \right) \;\;\mbox{where}\;-\infty < x< \infty \nonumber 
\end{eqnarray}
On the above,  $_1F_1( a,b,x)= \sum_{n=0}^{\infty }\frac{(a)_n}{(b)_n}\frac{x^n}{n!}$.

put $c_0= \left( -\frac{1}{2}\mu \right)^{1-\gamma } \frac{\Gamma (\psi _0+2-\gamma )}{\Gamma (2-\gamma )}$ as $\lambda = 1-\nu = 2(1-\gamma )$ in (\ref{eq:7}) with replacing $\beta _i$ by $\psi _i$.
\begin{rmk}
The power series expansion of GCH equation of the second kind for a polynomial which makes $B_n$ term terminated around $x=0 $ as $\Omega = -2\mu (\psi _i +1-\gamma +\frac{i}{2})$ where $i, \psi _i \in \mathbb{N}_{0}$ is given by
\begin{eqnarray}
 y(x)&=& RW_{\psi _i}\left( \psi _i=-\frac{\Omega }{2\mu }+\gamma -1-\frac{i}{2}, \omega, \gamma =\frac{1}{2}(1+\nu );\; \tilde{\varepsilon }= -\frac{1}{2}\varepsilon x;\; z=-\frac{1}{2}\mu x^2 \right) \nonumber\\
&&= z^{1-\gamma }\frac{\Gamma (\psi _0+2-\gamma )}{\Gamma (2-\gamma )} \Bigg\{\sum_{i_0=0}^{\psi _0} \frac{(-\psi _0)_{i_0}}{(1)_{i_0}(2-\gamma)_{i_0}}z^{i_0}\nonumber\\
&&+  \Bigg\{  \sum_{i_0=0}^{\psi _0}\frac{(i_0+1-\gamma +\frac{\omega }{2})}{(i_0+\frac{1}{2})(i_0+\frac{3}{2}-\gamma )} \frac{(-\psi _0)_{i_0}}{(1)_{i_0}(2-\gamma)_{i_0}}
  \sum_{i_1=i_0}^{\psi _1} \frac{(-\psi _1)_{i_1}(\frac{3}{2})_{i_0}(\frac{5}{2}-\gamma )_{i_0}}{(-\psi _1)_{i_0}(\frac{3}{2})_{i_1}(\frac{5}{2}-\gamma )_{i_1}} z^{i_1} \Bigg\} \tilde{\varepsilon }\nonumber\\
&&+ \sum_{n=2}^{\infty } \Bigg\{ \sum_{i_0=0}^{\psi _0} \frac{(i_0+1-\gamma +\frac{\omega }{2})}{(i_0+\frac{1}{2})(i_0+\frac{3}{2}-\gamma)} \frac{(-\psi _0)_{i_0}}{(1)_{i_0}(2-\gamma )_{i_0}} \nonumber\\
&&\times \prod _{k=1}^{n-1} \Bigg\{ \sum_{i_k=i_{k-1}}^{\psi _k} \frac{(i_k+1-\gamma +\frac{\omega }{2}+\frac{k}{2})}{(i_k+\frac{1}{2}+\frac{k}{2})(i_k+\frac{3}{2}-\gamma + \frac{k}{2})} \frac{(-\psi _k)_{i_k}(1+\frac{k}{2})_{i_{k-1}}(2-\gamma +\frac{k}{2})_{i_{k-1}}}{(-\psi _k)_{i_{k-1}}(1+\frac{k}{2})_{i_k}(2-\gamma +\frac{k}{2})_{i_k}}\Bigg\} \nonumber\\
&&\times \sum_{i_n= i_{n-1}}^{\psi _n} \frac{(-\psi _n)_{i_N}(1+\frac{n}{2})_{i_{n-1}}(2-\gamma +\frac{n}{2})_{i_{n-1}}}{(-\psi _n)_{i_{n-1}}(1+\frac{n}{2})_{i_n}(2-\gamma +\frac{n}{2})_{i_n}} z^{i_n}\Bigg\}\tilde{\varepsilon }^n \Bigg\}\nonumber
\end{eqnarray}
\end{rmk}
For the minimum value of the GCH equation of the second kind for a polynomial which makes $B_n$ term terminated around $x=0 $, put $\psi _0=\psi _1=\psi _2=\cdots=0$ in Remark 2.
\begin{eqnarray}
 y(x)&=& RW_{0}\left( \Omega = \mu (\nu -1-i), \omega, \gamma =\frac{1}{2}(1+\nu );\; \tilde{\varepsilon }= -\frac{1}{2}\varepsilon x;\; z=-\frac{1}{2}\mu x^2 \right) \nonumber\\
&=& z^{1-\gamma }\; _1F_1\left( \omega -\nu +1,2-\nu ,-\varepsilon x \right)  \;\;\mbox{where}\;-\infty < x< \infty \nonumber
\end{eqnarray} 
\subsection{Infinite series}
\begin{thm}
In Ref.\cite{Chou2012b}, the general expression of a power series of $y(x)$ for an infinite series is defined by
\begin{eqnarray}
y(x)  &=& \sum_{n=0}^{\infty } y_{n}(x)= y_0(x)+ y_1(x)+ y_2(x)+ y_3(x)+\cdots \nonumber\\
&=& c_0 \Bigg\{ \sum_{i_0=0}^{\infty } \left( \prod _{i_1=0}^{i_0-1}B_{2i_1+1} \right) x^{2i_0+\lambda } 
+ \sum_{i_0=0}^{\infty }\left\{ A_{2i_0} \prod _{i_1=0}^{i_0-1}B_{2i_1+1}  \sum_{i_2=i_0}^{\infty } \left( \prod _{i_3=i_0}^{i_2-1}B_{2i_3+2} \right)\right\} x^{2i_2+1+\lambda }  \nonumber\\
&& + \sum_{N=2}^{\infty } \Bigg\{ \sum_{i_0=0}^{\infty } \Bigg\{A_{2i_0}\prod _{i_1=0}^{i_0-1} B_{2i_1+1} 
 \prod _{k=1}^{N-1} \Bigg( \sum_{i_{2k}= i_{2(k-1)}}^{\infty } A_{2i_{2k}+k}\prod _{i_{2k+1}=i_{2(k-1)}}^{i_{2k}-1}B_{2i_{2k+1}+(k+1)}\Bigg)\nonumber\\
&& \times  \sum_{i_{2N} = i_{2(N-1)}}^{\infty } \Bigg( \prod _{i_{2N+1}=i_{2(N-1)}}^{i_{2N}-1} B_{2i_{2N+1}+(N+1)} \Bigg) \Bigg\} \Bigg\} x^{2i_{2N}+N+\lambda }\Bigg\} 
\label{eq:11}
\end{eqnarray}
\end{thm}
Substitute (\ref{eq:4a})-(\ref{eq:4c}) into (\ref{eq:11}). 
The general expression of a power series of GCH equation for an infinite series is given by
\begin{eqnarray}
 y(x) &=& \sum_{n=0}^{\infty } y_{n}(x)= y_0(x)+ y_1(x)+ y_2(x)+ y_3(x)+\cdots \nonumber\\
&=& c_0 x^{\lambda } \Bigg\{\sum_{i_0=0}^{\infty } \frac{(\frac{\Omega }{2\mu }+ \frac{\lambda }{2})_{i_0}}{(1+\frac{\lambda }{2})_{i_0}(\gamma +\frac{\lambda }{2})_{i_0}}z^{i_0} \nonumber\\
&&+   \Bigg\{ \sum_{i_0=0}^{\infty }\frac{(i_0+\frac{\lambda }{2}+\frac{\omega }{2})}{(i_0+\frac{1}{2}+\frac{\lambda }{2})(i_0-\frac{1}{2}+\gamma +\frac{\lambda }{2})} \frac{(\frac{\Omega }{2\mu }+\frac{\lambda }{2})_{i_0}}{(1+\frac{\lambda }{2})_{i_0}(\gamma +\frac{\lambda }{2})_{i_0}}\nonumber\\
&&\times  \sum_{i_1=i_0}^{\infty }\frac{(\frac{\Omega }{2\mu }+\frac{1}{2}+\frac{\lambda }{2})_{i_1}(\frac{3}{2}+\frac{\lambda }{2})_{i_0}(\gamma +\frac{1}{2}+ \frac{\lambda }{2})_{i_0}}{(\frac{\Omega }{2\mu }+\frac{1}{2}+\frac{\lambda }{2})_{i_0}(\frac{3}{2}+\frac{\lambda }{2})_{i_1}(\gamma +\frac{1}{2}+\frac{\lambda }{2})_{i_1}} z^{i_1} \Bigg\} \tilde{\varepsilon }\nonumber\\
&&+ \sum_{n=2}^{\infty } \Bigg\{ \sum_{i_0=0}^{\infty } \frac{(i_0+\frac{\lambda }{2}+\frac{\omega }{2})}{(i_0+\frac{1}{2}+\frac{\lambda }{2})(i_0-\frac{1}{2}+\gamma +\frac{\lambda }{2})} \frac{(\frac{\Omega }{2\mu }+\frac{\lambda }{2})_{i_0}}{(1+\frac{\lambda }{2})_{i_0}(\gamma +\frac{\lambda }{2})_{i_0}} \nonumber\\
&&\times \prod _{k=1}^{n-1} \Bigg\{ \sum_{i_k=i_{k-1}}^{\infty } \frac{(i_k+\frac{\lambda }{2}+\frac{\omega }{2}+\frac{k}{2})}{(i_k+\frac{1}{2}+\frac{\lambda }{2}+\frac{k}{2})(i_k-\frac{1}{2}+\gamma + \frac{k}{2}+\frac{\lambda }{2})} \nonumber\\
&&\times \frac{(\frac{\Omega }{2\mu }+\frac{k}{2}+\frac{\lambda }{2})_{i_k}(1+\frac{k}{2}+\frac{\lambda }{2})_{i_{k-1}}(\frac{k}{2}+\gamma +\frac{\lambda }{2})_{i_{k-1}}}{(\frac{\Omega }{2\mu }+\frac{k}{2}+\frac{\lambda }{2})_{i_{k-1}}(1+\frac{k}{2}+\frac{\lambda }{2})_{i_k}(\frac{k}{2}+\gamma +\frac{\lambda }{2})_{i_k}}\Bigg\} \nonumber\\
&&\times \sum_{i_n= i_{n-1}}^{\infty } \frac{(\frac{\Omega }{2\mu }+\frac{n}{2}+\frac{\lambda }{2})_{i_n}(1+\frac{n}{2}+\frac{\lambda }{2})_{i_{n-1}}(\frac{n}{2}+\gamma +\frac{\lambda }{2})_{i_{n-1}}}{(\frac{\Omega }{2\mu }+\frac{n}{2}+\frac{\lambda }{2})_{i_{n-1}}(1+\frac{n}{2}+\frac{\lambda }{2})_{i_n}(\frac{n}{2}+\gamma +\frac{\lambda }{2})_{i_n}} z^{i_n}\Bigg\} \tilde{\varepsilon }^n \Bigg\}
\label{eq:12}
\end{eqnarray}
Put $c_0$= $\frac{\Gamma (\gamma -\frac{\Omega }{2\mu })}{\Gamma (\gamma )}$ as $\lambda =0$ for the first independent solution of GCH equation and $c_0= \left( -\frac{1}{2}\mu \right)^{1-\gamma } \frac{\Gamma (1-\frac{\Omega }{2\mu })}{\Gamma (2-\gamma )}$ as $\lambda = 1-\nu = 2(1-\gamma )$ for the second one in (\ref{eq:12})
\begin{rmk}
The power series expansion of GCH equation of the first kind for an infinite series around $x=0 $ is given by
\begin{eqnarray}
 y(x)&=& QW\left(\omega, \gamma =\frac{1}{2}(1+\nu );\; \tilde{\varepsilon }= -\frac{1}{2}\varepsilon x;\; z=-\frac{1}{2}\mu x^2 \right) \nonumber\\
&&= \frac{\Gamma (\gamma -\frac{\Omega }{2\mu })}{\Gamma (\gamma )} \Bigg\{\sum_{i_0=0}^{\infty } \frac{(\frac{\Omega }{2\mu })_{i_0}}{(1)_{i_0}(\gamma)_{i_0}}z^{i_0}+  \Bigg\{\sum_{i_0=0}^{\infty } \frac{(i_0+\frac{\omega }{2})}{(i_0+\frac{1}{2})(i_0-\frac{1}{2}+\gamma )} \frac{(\frac{\Omega }{2\mu })_{i_0}}{(1)_{i_0}(\gamma)_{i_0}}\nonumber\\
&&\times  \sum_{i_1=i_0}^{\infty }  \frac{(\frac{\Omega }{2\mu }+\frac{1}{2})_{i_1}(\frac{3}{2})_{i_0}(\gamma +\frac{1}{2})_{i_0}}{(\frac{\Omega }{2\mu }+\frac{1}{2})_{i_0}(\frac{3}{2})_{i_1}(\gamma +\frac{1}{2})_{i_1}} z^{i_1} \Bigg\} \tilde{\varepsilon } 
+ \sum_{n=2}^{\infty } \Bigg\{ \sum_{i_0=0}^{\infty } \frac{(i_0+\frac{\omega }{2})}{(i_0+\frac{1}{2})(i_0-\frac{1}{2}+\gamma)} \frac{(\frac{\Omega }{2\mu })_{i_0}}{(1)_{i_0}(\gamma )_{i_0}} \nonumber\\
&&\times \prod _{k=1}^{n-1} \Bigg\{ \sum_{i_k=i_{k-1}}^{\infty } \frac{(i_k+\frac{\omega }{2}+\frac{k}{2})}{(i_k+\frac{1}{2}+\frac{k}{2})(i_k-\frac{1}{2}+\gamma + \frac{k}{2})} \frac{(\frac{\Omega }{2\mu }+\frac{k}{2})_{i_k}(1+\frac{k}{2})_{i_{k-1}}(\frac{k}{2}+\gamma )_{i_{k-1}}}{(\frac{\Omega }{2\mu }+\frac{k}{2})_{i_{k-1}}(1+\frac{k}{2})_{i_k}(\frac{k}{2}+\gamma )_{i_k}}\Bigg\} \nonumber\\
&&\times \sum_{i_n= i_{n-1}}^{\infty } \frac{(\frac{\Omega }{2\mu }+\frac{n}{2})_{i_n}(1+\frac{n}{2})_{i_{n-1}}(\frac{n}{2}+\gamma )_{i_{n-1}}}{(\frac{\Omega }{2\mu }+\frac{n}{2})_{i_{n-1}}(1+\frac{n}{2})_{i_n}(\frac{n}{2}+\gamma )_{i_n}} z^{i_n}\Bigg\} \tilde{\varepsilon }^n \Bigg\}\nonumber
\end{eqnarray}
\end{rmk}
\begin{rmk}
The power series expansion of GCH equation of the second kind for an infinite series around $x=0 $ is given by
\begin{eqnarray}
 y(x)&=& RW\left(\omega, \gamma =\frac{1}{2}(1+\nu );\; \tilde{\varepsilon }= -\frac{1}{2}\varepsilon x;\; z=-\frac{1}{2}\mu x^2 \right) \nonumber\\
&&= z^{1-\gamma }\frac{\Gamma (1-\frac{\Omega }{2\mu })}{\Gamma (2-\gamma )} \Bigg\{\sum_{i_0=0}^{\infty } \frac{(\frac{\Omega }{2\mu }+1-\gamma )_{i_0}}{(1)_{i_0}(2-\gamma)_{i_0}}z^{i_0} \nonumber\\
&&+  \Bigg\{ \sum_{i_0=0}^{\infty }\frac{(i_0+1-\gamma +\frac{\omega }{2})}{(i_0+\frac{1}{2})(i_0+\frac{3}{2}-\gamma )} \frac{(\frac{\Omega }{2\mu }+1-\gamma )_{i_0}}{(1)_{i_0}(2-\gamma)_{i_0}}  \sum_{i_1=i_0}^{\infty } \frac{(\frac{\Omega }{2\mu }+\frac{3}{2}-\gamma )_{i_1}(\frac{3}{2})_{i_0}(\frac{5}{2}-\gamma )_{i_0}}{(\frac{\Omega }{2\mu }+\frac{3}{2}-\gamma )_{i_0}(\frac{3}{2})_{i_1}(\frac{5}{2}-\gamma )_{i_1}} z^{i_1} \Bigg\} \tilde{\varepsilon } \nonumber\\
&&+ \sum_{n=2}^{\infty } \Bigg\{ \sum_{i_0=0}^{\infty }\frac{(i_0+1-\gamma +\frac{\omega }{2})}{(i_0+\frac{1}{2})(i_0+\frac{3}{2}-\gamma)} \frac{(\frac{\Omega }{2\mu }+1-\gamma )_{i_0}}{(1)_{i_0}(2-\gamma )_{i_0}}\nonumber\\
&&\times  \prod _{k=1}^{n-1} \Bigg\{ \sum_{i_k=i_{k-1}}^{\infty } \frac{(i_k+1-\gamma +\frac{\omega }{2}+\frac{k}{2})}{(i_k+\frac{1}{2}+\frac{k}{2})(i_k+\frac{3}{2}-\gamma + \frac{k}{2})} \nonumber\\
&&\times \frac{(\frac{\Omega }{2\mu }+1-\gamma + \frac{k}{2})_{i_k}(1+\frac{k}{2})_{i_{k-1}}(2-\gamma +\frac{k}{2})_{i_{k-1}}}{(\frac{\Omega }{2\mu }+1-\gamma + \frac{k}{2})_{i_{k-1}}(1+\frac{k}{2})_{i_k}(2-\gamma +\frac{k}{2})_{i_k}}\Bigg\} \nonumber\\
&&\times \sum_{i_n= i_{n-1}}^{\infty } \frac{(\frac{\Omega }{2\mu }+1-\gamma + \frac{n}{2})_{i_n}(1+\frac{n}{2})_{i_{n-1}}(2-\gamma +\frac{n}{2})_{i_{n-1}}}{(\frac{\Omega }{2\mu }+1-\gamma + \frac{n}{2})_{i_{n-1}}(1+\frac{n}{2})_{i_n}(2-\gamma +\frac{n}{2})_{i_n}} z^{i_n}\Bigg\} \tilde{\varepsilon }^n \Bigg\}\nonumber
\end{eqnarray}
\end{rmk}

 When $\nu $ is integer, one of two solution of the GCH equation does not have any meaning, because $RW_{\psi _i}\left( \psi _i=-\frac{\Omega }{2\mu }+\gamma -1-\frac{i}{2}, \gamma =\frac{1}{2}(1+\nu );\; \tilde{\varepsilon };\; z \right)$ can be described as $QW_{\beta _i}\left( \beta _i=-\frac{\Omega }{2\mu }-\frac{i}{2} , \gamma =\frac{1}{2}(1+\nu );\; \tilde{\varepsilon };\; z\right)$ as long as $|\lambda _1-\lambda _2|=|\nu -1|=$ integer. As we see remarks 2.2, 2.3, 2.5 and 2.6, it is required that $\nu \ne 0,-1,-2,\cdots$ for the first kind of independent solutions of GCH equation for a polynomial which makes $B_n$ term terminated and an infinite series. By similar reason,  $\nu \ne 2,3,4,\cdots$ is required for the second kind of independent solutions of GCH equation. 
\section{Asymptotic behavior of the function $\bm{y(x)}$ and the boundary condition for $\bm{x}$}
\subsection{The case of $|\mu|  \ll 1$ or $ |\mu| \ll |\varepsilon | $}
As $n\gg 1$ (for sufficiently large $n$), (\ref{eq:3})--(\ref{eq:4b}) are asymptotically equal to
\begin{subequations}
\begin{equation}
c_{n+1}=A\;c_n +B\;c_{n-1} \hspace{1cm};n\geq 1 \label{eq:100}
\end{equation}
where
\begin{equation}
\lim_{n\gg 1} A_n = A= -\frac{\varepsilon }{n}
\label{eq:37a}
\end{equation}
And,
\begin{equation}
\lim_{n\gg 1} B_n = B= -\frac{\mu }{n}
\label{eq:37b}
\end{equation}
\end{subequations}
With $c_1\sim  A\;c_0$.\footnote{We only have the sense of curiosity about an asymptotic series as  $n\gg 1$ for given $x$. Actually, $c_1 =  A_0 c_0$. But for a huge value of an index $n$, I treat the coefficient $c_1$ as $ A c_0$ for simple computations.} Since $|\mu|  \ll 1$ or $ |\mu| \ll |\varepsilon | $, (\ref{eq:37b}) is negligible.
Its recurrence relation is
\begin{equation}
c_{n+1} = -\frac{\varepsilon }{n} c_n
\label{eq:39}
\end{equation}
Plug (\ref{eq:39}) into the power series expansion where $ \sum_{n=0}^{\infty } c_n x^n $, putting $c_0= -0/\varepsilon  $ for simplicity.
\begin{equation}
\lim_{n\gg 1}y(x) =  x e^{-\varepsilon x} \hspace{1cm}\mbox{where}\;-\infty <x< \infty 
\label{eq:40}
\end{equation}
\subsection{The case of $|\varepsilon |  \ll 1$ or $ |\varepsilon | \ll |\mu | $}
Let assume that $|\varepsilon |  \ll 1$ or $ |\varepsilon | \ll |\mu | $. Then (\ref{eq:37a}) is negligible.
Its recurrence relation is
\begin{equation}
c_{n+1} = -\frac{\mu }{n} c_{n-1}
\label{eq:42}
\end{equation}
We can classify $c_n$ as to even and odd terms in (\ref{eq:42}).
\begin{equation}
c_{2n} = \frac{\big(-\frac{1}{2}\big)!}{\big(n-\frac{1}{2}\big)!}\Big( -\frac{1}{2}\mu \Big)^n c_0 \hspace{1cm}c_{2n+1} = \frac{1 }{n!}\Big( -\frac{1}{2}\mu \Big)^n c_1\hspace{1cm}\mbox{where}\;n\geq 1
\label{eq:43}
\end{equation}
$c_1\sim  A c_0 =0$ in (\ref{eq:43}). Because $A$ is negligible since $|\varepsilon |  \ll 1$ or $ |\varepsilon | \ll |\mu | $. 
Put $c_{2n}$ in (\ref{eq:43}) into the power series expansion where $ \sum_{n=0}^{\infty } c_n x^n $, putting $c_0 =1$ for simplicity.
\begin{eqnarray}
\lim_{n\gg 1}y(x) &=& 1+  \sqrt{ -\frac{\pi }{2}\mu x^2} \mbox{Erf}\left(\sqrt{ -\frac{1}{2}\mu x^2}\right) e^{ -\frac{1}{2}\mu x^2} \label{eq:44}\\
&&\mbox{where}\;-\infty <x< \infty \nonumber
\end{eqnarray} 
On the above $\mbox{Erf(y)} $ is an error function which is
\begin{equation}
\mbox{Erf(y)} = \frac{2}{\sqrt{\pi }} \int_{0}^{y} dt\; e^{-t^2}\nonumber
\end{equation} 

\section{Application}
I show power series expansions in closed forms and asymptotic behaviors of the GCH function in this paper. 
We can apply this new special functions into many physics areas. I show three examples of GCH equation as follows:
\subsection{the rotating harmonic oscillator}
For example, there are quantum-mechanical systems whose radial Schr$\ddot{\mbox{o}}$dinger equation may be reduced to a Biconfluent Heun function\cite{Leau1986,Mass1983}, namely the rotating harmonic oscillator and a class of confinement potentials. Its radial Schr$\ddot{\mbox{o}}$dinger equation is given by
\begin{equation}
\Psi^{''}(r)+ \bigg\{ \frac{2\lambda _m+1}{2\omega } -\frac{(r-1)^2}{4\omega ^2}- \frac{l_m (l_m+1)}{r^2}\bigg\} \Psi(r) =0  
\label{eq:47}
\end{equation}
where $0\leq  r < \infty $, ¸$\lambda _m$ is the eigenvalue, $l_m$ is the rotational quantum number and $\omega$ is a
coupling parameter.
By means of the changes of variable,
\begin{equation}
\Psi(r) = r^{l_m+1} \exp\left( -\frac{(r-1)^2}{2\omega } \right) U(r) \hspace{.5cm}\mbox{and}\hspace{.5cm}  r=\sqrt{2\omega }x
\label{eq:48}
\end{equation}
the above becomes the following Biconfluent Heun equation:
\begin{equation}
x U^{''}(x)+ (1+\alpha -\beta x -2 x^2)U^{'}(x)+ \left\{ (\gamma -\alpha -2)x - \frac{1}{2}[\delta +\beta (1+\alpha )] \right\} U(x) =0  
\label{eq:49}
\end{equation}
where the four Heun parameters are
\begin{equation}
\alpha =2l_m+1 \hspace{.5cm}\beta = -\sqrt{\frac{2}{\omega }}\hspace{.5cm} \delta =0 \hspace{.5cm}\gamma = 1+2\lambda_m 
\label{eq:50}
\end{equation}
If we compare (\ref{eq:49}) with (\ref{eq:1}), all coefficients on the above are correspondent to the following way.
\begin{equation}
\begin{split}
& \mu  \longleftrightarrow   -2 \\ & \varepsilon  \longleftrightarrow  -\beta  \\ & \nu  \longleftrightarrow  1+\alpha  \\
& \Omega  \longleftrightarrow  \gamma -\alpha -2  \\ & \omega  \longleftrightarrow \frac{1}{2\beta }[\delta +\beta (1+\alpha )]
\end{split}\label{eq:56}   
\end{equation}
Put (\ref{eq:50}) in (\ref{eq:56}).
\begin{equation}
\begin{split}
& \mu  \longleftrightarrow   -2 \\ & \varepsilon  \longleftrightarrow  \sqrt{\frac{2}{\omega }} \\ & \nu  \longleftrightarrow  2(l_m+1)  \\
& \Omega  \longleftrightarrow  2(\lambda _m-l_m-1)  \\ & \omega  \longleftrightarrow l_m+1
\end{split}\label{eq:57}   
\end{equation}
Let's investigate function $\Psi(r)$ as $n$ and $r$ go to infinity. I assume that $U(x)$ is infinite series in (\ref{eq:49}). Since $\varepsilon \ll 1$ in (\ref{eq:57}), put (\ref{eq:44}) in (\ref{eq:48}) with replacing $\mu $ by $-2$.
\begin{equation}
\lim_{n\gg 1}\Psi(r) \approx r^{l_m+1} \exp\left(-\frac{(r-1)^2}{2\omega }\right) \left( 1+  \sqrt{\frac{\pi}{2\omega}} \mbox{Erf}\left(\frac{r}{\sqrt{ 2\omega }} \right) r e^{\frac{r^2}{2\omega}}\right) 
\label{eq:58}
\end{equation}
In  (\ref{eq:58}) if $r\rightarrow \infty $, then $\displaystyle {\lim_{n\gg 1}\Psi(r)\rightarrow \infty }$. It is unacceptable that wave function $\Psi(r)$ is divergent as $r$ goes to infinity in the quantum mechanical point of view. Therefore the function $U(x)$ must to be polynomial in (\ref{eq:49}) in order to make the wave function $\Psi(r)$ being convergent even if $r$ goes to infinity. $RW_{\psi _i}\left( \psi _i, \omega, \gamma;\; \tilde{\varepsilon }= -\frac{r}{2\omega};\; z=\frac{r^2}{2\omega} \right)\rightarrow \infty $ as $r\rightarrow 0$ because of $ \gamma =l_m+ \frac{3}{2}$ in Remark 2.3. But $QW_{\beta _i}\left( \beta _i,\omega, \gamma ;\; \tilde{\varepsilon }= -\frac{r}{2\omega};\; z=\frac{r^2}{2\omega} \right)\rightarrow 0$ as $r\rightarrow 0$ in Remark 2.2. So I choose Remark 2.2 as eigenfunction for (\ref{eq:48}). Put (\ref{eq:57}) in Remark 2.2 replacing $x$ and $y(x)$ by $\frac{r}{\sqrt{2\omega}}$ and $U(r)$.   
\begin{eqnarray}
  U(r)&=& QW_{\beta _i}\left( \beta _i=\frac{\lambda _m-l_m-1-i}{2} ,\omega =l_m+1, \gamma =l_m+ \frac{3}{2};\; \tilde{\varepsilon }= -\frac{r}{2\omega};\; z=\frac{r^2}{2\omega} \right)\nonumber\\
&&= \frac{\Gamma (\gamma +\beta _0)}{\Gamma (\gamma )} \Bigg\{\sum_{i_0=0}^{\beta _0 } \frac{(-\beta _0)_{i_0}}{(1)_{i_0}(\gamma)_{i_0}}z^{i_0}+   \Bigg\{\sum_{i_0=0}^{\beta _0 } \frac{(i_0+\frac{\omega }{2})}{(i_0+\frac{1}{2})(i_0-\frac{1}{2}+\gamma )} \frac{(-\beta  _0)_{i_0}}{(1)_{i_0}(\gamma)_{i_0}}\nonumber\\
&&\times  \sum_{i_1=i_0}^{\beta _1}\frac{(-\beta _1)_{i_1}(\frac{3}{2})_{i_0}(\gamma +\frac{1}{2})_{i_0}}{(-\beta _1)_{i_0}(\frac{3}{2})_{i_1}(\gamma +\frac{1}{2})_{i_1}} z^{i_1} \Bigg\} \tilde{\varepsilon } 
+ \sum_{n=2}^{\infty } \Bigg\{ \sum_{i_0=0}^{\beta _0} \frac{(i_0+\frac{\omega }{2})}{(i_0+\frac{1}{2})(i_0-\frac{1}{2}+\gamma)} \frac{(-\beta _0)_{i_0}}{(1)_{i_0}(\gamma )_{i_0}} \nonumber\\
&&\times \prod _{k=1}^{n-1} \Bigg\{ \sum_{i_k=i_{k-1}}^{\beta _k} \frac{(i_k+\frac{\omega }{2}+\frac{k}{2})}{(i_k+\frac{1}{2}+\frac{k}{2})(i_k-\frac{1}{2}+\gamma + \frac{k}{2})} \frac{(-\beta _k)_{i_k}(1+\frac{k}{2})_{i_{k-1}}(\frac{k}{2}+\gamma )_{i_{k-1}}}{(-\beta _k)_{i_{k-1}}(1+\frac{k}{2})_{i_k}(\frac{k}{2}+\gamma )_{i_k}}\Bigg\} \nonumber\\
&&\times \sum_{i_n= i_{n-1}}^{\beta _n} \frac{(-\beta _n)_{i_n}(1+\frac{n}{2})_{i_{n-1}}(\frac{n}{2}+\gamma )_{i_{n-1}}}{(-\beta _n)_{i_{n-1}}(1+\frac{n}{2})_{i_n}(\frac{n}{2}+\gamma )_{i_n}} z^{i_n}\Bigg\} \tilde{\varepsilon }^n \Bigg\}
\label{eq:59}
\end{eqnarray}
Put (\ref{eq:59}) in (\ref{eq:48}). The wave function for the rotating harmonic oscillator is given by
\begin{eqnarray}
\Psi(r) &=& N r^{l_m+1} \exp\left(-\frac{(r-1)^2}{2\omega }\right) QW_{\beta _i}\left( \beta _i=\frac{\lambda _m-l_m-1-i}{2} ,\omega =l_m+1, \gamma =l_m+ \frac{3}{2}\right.\nonumber\\
&&;\left. \tilde{\varepsilon }= -\frac{r}{2\omega};\; z=\frac{r^2}{2\omega} \right)
\label{eq:60}
\end{eqnarray}
N is normalized constant. Eigenvalue $\lambda _m$ is 
\begin{equation}
\lambda _m  =2\beta _i+l_m+1+i \hspace{1cm} \mathrm{where}\; i,\beta _i \in \mathbb{N}_{0} \nonumber
\end{equation}
In general most of well-known special function with two recursive coefficients  (Bessel, Legendre, Kummer, Laguerre and hypergeometric functions, etc) only has one eigenvalue for the polynomial case. However the GCH function with three recursive coefficients has infinite eigenvalues as we see (\ref{eq:59}).
\subsection{Confinement potentials}
Following Chaudhuri and Mukherjee, there is the radial Schr$\ddot{\mbox{o}}$dinger equation.\cite{Leau1986,Chau1983,Chau1984}:
\begin{equation}
\Psi^{''}(r)+ \bigg\{ \bigg( \frac{2\mu }{\hbar ^2} \bigg) \bigg( E+ \frac{a}{r} -br- cr^2\bigg) -\frac{l(l+1)}{r^2} \bigg\} \Psi(r) =0  
\label{eq:51}
\end{equation}
with $E$ being the energy. By means of the consecutive changes of variable
\begin{equation}
\Psi(r) = r^{l+1} \exp\left( -\frac{1}{2} \alpha _F r^2 -\beta _F r \right) U(r) \hspace{.5cm}\mbox{and}\hspace{.5cm} x=\sqrt{\alpha _F}r
\label{eq:52}
\end{equation}
the above becomes also the following Biconfluent Heun equation:
\begin{equation}
x U^{''}(x)+ (1+\alpha -\beta x -2 x^2)U^{'}(x)+ \left\{ (\gamma -\alpha -2)x - \frac{1}{2}[\delta +\beta (1+\alpha )] \right\} U(x) =0  
\label{eq:53}
\end{equation}
where the four Heun parameters are
\begin{eqnarray}
&&\alpha = 2l+1,\hspace{.5cm} \gamma = \frac{\epsilon _F}{\alpha _F},\nonumber\\
&& \beta = 2\frac{\beta _F}{\sqrt{\alpha _F}},\hspace{.5cm} \delta = -\frac{4\mu }{\hbar ^2}\frac{a}{\sqrt{\alpha _F}}
\label{eq:54}
\end{eqnarray}
where,
\begin{equation}
\alpha_F = \bigg( \frac{2\mu c}{\hbar ^2} \bigg)^{1/2} , \hspace{.5cm} \beta _F = b\bigg( \frac{ \mu }{2\hbar ^2 c}\bigg)^{1/2} , \hspace{.5cm}\epsilon _F = \beta _F^2 +\frac{2\mu }{\hbar ^2} E
\label{eq:55}
\end{equation}
Put (\ref{eq:54}) and (\ref{eq:55}) in (\ref{eq:56}).
\begin{equation}
\begin{split}
& \mu  \longleftrightarrow   -2 \\ & \varepsilon  \longleftrightarrow  -2\frac{\beta _F}{\sqrt{\alpha _F}} \\ & \nu  \longleftrightarrow  2(l+1) \\
& \Omega  \longleftrightarrow   \frac{\epsilon _F}{ \alpha _F } -2\left( l+\frac{3}{2}\right) =  \frac{1}{ \alpha _F }\left( \beta _F^2 +\frac{2\mu }{\hbar ^2} E\right) -2\left( l+\frac{3}{2}\right) \\ & \omega  \longleftrightarrow -\frac{\mu a }{\hbar ^2 \beta _F}+ l+1 
\end{split}\label{eq:61}   
\end{equation}
Let's investigate function $\Psi(r)$ as $n$ and $r$ go to infinity. I assume that $U(x)$ is infinite series in (\ref{eq:53}). Since $ |\varepsilon |\ll 1 $ in (\ref{eq:61}), put (\ref{eq:44}) in (\ref{eq:52}) with replacing $\mu $ and $x$ by $-2$ and $\sqrt{\alpha _F}r$.
\begin{equation}
\lim_{n\gg 1}\Psi(r) \approx r^{l+1} \exp\left(-\frac{\alpha _F}{2}r^2-\beta _F r\right) \left( 1+  \sqrt{\pi \alpha _F}  \mbox{Erf}\left(\sqrt{ \alpha _F }r \right) r  e^{\alpha _F r^2}\right) 
\label{eq:62}
\end{equation}
In  (\ref{eq:62}) if $r\rightarrow \infty $, then $\displaystyle {\lim_{n\gg 1}\Psi(r)\rightarrow \infty }$. It is unacceptable that wave function $\Psi(r)$ is divergent as $r$ goes to infinity in the quantum mechanical point of view. Therefore the function $U(x)$ must to be polynomial in (\ref{eq:53}) in order to make the wave function $\Psi(r)$ being convergent even if $r$ goes to infinity. $RW_{\psi _i}\left( \psi _i, \omega, \gamma;\; \tilde{\varepsilon }= -\beta _F r;\; z= \alpha _F r^2 \right)\rightarrow \infty $ as $r\rightarrow 0$ because of $ \gamma = l+\frac{3}{2}$ in Remark 2.3. But $QW_{\beta _i}\left( \beta _i,\omega, \gamma ;\; \tilde{\varepsilon }= -\beta _F r;\; z=\alpha _F r^2 \right)\rightarrow 0$ as $r\rightarrow 0$ in Remark 2.2. So I choose Remark 2.2 as eigenfunction for (\ref{eq:52}). Put (\ref{eq:61}) in Remark 2.2 replacing $x$ and $y(x)$ by $\sqrt{\alpha _F}r$ and $U(r)$.  
\begin{eqnarray}
 U(r)&=& QW_{\beta _i}\left( \beta _i= \frac{1}{4 \alpha _F }\left( \beta _F^2 +\frac{2\mu }{\hbar ^2}E \right)-\frac{1}{2}\left(i+l+\frac{3}{2} \right),\omega =-\frac{\mu a }{\hbar ^2 \beta _F}+ l+1 \right.\nonumber\\
&&,\left. \gamma = l+ \frac{3}{2};\; \tilde{\varepsilon }= -\beta _F r;\; z=\alpha _F r^2\right) \nonumber\\
&&= \frac{\Gamma (\gamma +\beta _0)}{\Gamma (\gamma )} \Bigg\{\sum_{i_0=0}^{\beta _0 } \frac{(-\beta _0)_{i_0}}{(1)_{i_0}(\gamma)_{i_0}}z^{i_0}+   \Bigg\{ \sum_{i_0=0}^{\beta _0 }\frac{(i_0+\frac{\omega }{2})}{(i_0+\frac{1}{2})(i_0-\frac{1}{2}+\gamma )} \frac{(-\beta  _0)_{i_0}}{(1)_{i_0}(\gamma)_{i_0}}\nonumber\\
&&\times  \sum_{i_1=i_0}^{\beta _1} \frac{(-\beta _1)_{i_1}(\frac{3}{2})_{i_0}(\gamma +\frac{1}{2})_{i_0}}{(-\beta _1)_{i_0}(\frac{3}{2})_{i_1}(\gamma +\frac{1}{2})_{i_1}} z^{i_1} \Bigg\}\tilde{\varepsilon } 
+ \sum_{n=2}^{\infty } \Bigg\{ \sum_{i_0=0}^{\beta _0} \frac{(i_0+\frac{\omega }{2})}{(i_0+\frac{1}{2})(i_0-\frac{1}{2}+\gamma)} \frac{(-\beta _0)_{i_0}}{(1)_{i_0}(\gamma )_{i_0}} \nonumber\\
&&\times \prod _{k=1}^{n-1} \Bigg\{ \sum_{i_k=i_{k-1}}^{\beta _k} \frac{(i_k+\frac{\omega }{2}+\frac{k}{2})}{(i_k+\frac{1}{2}+\frac{k}{2})(i_k-\frac{1}{2}+\gamma + \frac{k}{2})} \frac{(-\beta _k)_{i_k}(1+\frac{k}{2})_{i_{k-1}}(\frac{k}{2}+\gamma )_{i_{k-1}}}{(-\beta _k)_{i_{k-1}}(1+\frac{k}{2})_{i_k}(\frac{k}{2}+\gamma )_{i_k}}\Bigg\} \nonumber\\
&&\times \sum_{i_n= i_{n-1}}^{\beta _n} \frac{(-\beta _n)_{i_n}(1+\frac{n}{2})_{i_{n-1}}(\frac{n}{2}+\gamma )_{i_{n-1}}}{(-\beta _n)_{i_{n-1}}(1+\frac{n}{2})_{i_n}(\frac{n}{2}+\gamma )_{i_n}} z^{i_n}\Bigg\} \tilde{\varepsilon }^n \Bigg\}
\label{eq:63}
\end{eqnarray}
Put (\ref{eq:63}) in (\ref{eq:52}). The wave function for confinement potentials is given by
\begin{eqnarray}
\Psi(r) &=& N r^{l+1} \exp\left( -\frac{1}{2}r^2\alpha _F -\beta _F r \right) QW_{\beta _i}\left( \beta _i= \frac{1}{4 \alpha _F }\left( \beta _F^2 +\frac{2\mu }{\hbar ^2}E \right)-\frac{1}{2}\left(i+l+\frac{3}{2} \right) \right.\nonumber\\
&&,\left. \omega =-\frac{\mu a }{\hbar ^2 \beta _F}+ l+1, \gamma = l+ \frac{3}{2};\; \tilde{\varepsilon }= -\beta _F r;\; z=\alpha _F r^2\right) \label{eq:65}
\end{eqnarray}
N is normalized constant. Energy $E$ is 
\begin{equation}
E= \frac{\hbar ^2}{2\mu } \left(  4 \alpha _F \left( \beta _i+\frac{i+l+\frac{3}{2}}{2} \right) -\beta _F^2\right)\hspace{1cm} \mathrm{where}\; i,\beta _i \in \mathbb{N}_{0} \nonumber
\end{equation}
Again the GCH function with three recursive coefficients has infinite eigenvalues.
\subsection{ The spin free Hamiltonian involving only scalar potential for the $q-\bar{q}$ system }
Following G\"{u}rsey and his colleagues, there is the spin free Hamiltonian involving only scalar potential for the $q-\bar{q}$ system:\cite{2011,1985,1988,1991}
\begin{equation} 
H^2 = 4\left[(m+ \frac{1}{2}br)^2 + P_r^2 + \frac{l(l+1)}{r^2}\right] 
\label{eq:67}
\end{equation}
where $P_r^2 = - \frac{\partial ^2}{\partial r^2} - \frac{2}{r} \frac{\partial}{\partial r} $, $m$= mass, $b$= real positive, and $l$= angular momentum quantum number. When wave function $\Psi (r) = \exp\left(-\frac{b}{4}\left(r+\frac{2m}{b}\right)^2\right) r^l y(r)Y_l^{m^{\star}}(\theta ,\phi)$ acts on both sides of (\ref{eq:67}), it becomes
\begin{equation}
r\frac{\partial^2{y}}{\partial{r}^2} + \left( - b r^2 -2m r +2(l+1)\right) \frac{\partial{y}}{\partial{r}} + \left( \left(\frac{E^2}{4}- b\left(l+\frac{3}{2}\right)\right) r  -2m(l+1)\right) y = 0
\label{eq:68}
\end{equation}
If we compare (\ref{eq:68}) with (\ref{eq:1}), all coefficients on the above are correspondent to the following way.
\begin{equation}
\begin{split}
& \mu  \longleftrightarrow   -b \\ & \varepsilon  \longleftrightarrow  -2m  \\ & \nu  \longleftrightarrow  2(l+1)  \\
& \Omega  \longleftrightarrow  \frac{E^2}{4}-b\left( l+\frac{3}{2}\right)  \\ & \omega  \longleftrightarrow l+1
\end{split}\label{eq:69}   
\end{equation}
Let's investigate function $\Psi(r)$ as $n$ and $r$ go to infinity. I assume that $y(r)$ is infinite series in (\ref{eq:68}). Since $\varepsilon \ll 1$ in (\ref{eq:69}), put (\ref{eq:44}) in $\Psi (r) = \exp\left(-\frac{b}{4}\left(r+\frac{2m}{b}\right)^2\right) r^l y(r)Y_l^{m^{\star}}(\theta ,\phi)$ with replacing $x$ and $\mu $ by $r$ and $-b$ .
\begin{equation}
\lim_{n\gg 1}\Psi(r) \approx r^l \exp\left(-\frac{b}{4}\left(r+\frac{2m}{b}\right)^2\right) \left\{ 1+ \sqrt{ \frac{1 }{2}b\pi} \mbox{Erf}\left(\sqrt{\frac{1}{2}b r^2}\right) r e^{ \frac{1}{2}b r^2}\right\}Y_l^{m^{\star}}(\theta ,\phi)
\label{eq:70}
\end{equation}
In  (\ref{eq:70}) if $r\rightarrow \infty $, then $\displaystyle {\lim_{n\gg 1}\Psi(r)\rightarrow \infty }$. It is unacceptable that wave function $\Psi(r)$ is divergent as $r$ goes to infinity in the quantum mechanical point of view. Therefore the function $y(r)$ must to be polynomial in (\ref{eq:68}) in order to make the wave function $\Psi(r)$ being convergent even if $r$ goes to infinity. $RW_{\psi _i}\left( \psi _i, \omega, \gamma;\; \tilde{\varepsilon }= mr;\; z=\frac{b}{2}\mu r^2\right)\rightarrow \infty $ as $r\rightarrow 0$ because of $\gamma = l+\frac{3}{2}$ in Remark 2.3. But $QW_{\beta _i}\left( \beta _i,\omega, \gamma ;\; \tilde{\varepsilon }= mr;\; z=\frac{b}{2}\mu r^2\right)\rightarrow 0$ as $r\rightarrow 0$ in Remark 2.2. So I choose Remark 2.2 as eigenfunction for (\ref{eq:68}). Put (\ref{eq:69}) in Remark 2.2 with replacing $x$ by $r$.
\begin{eqnarray}
 y(r)&=& QW_{\beta _i}\left( \beta _i= \frac{1}{2}\left( \frac{E^2}{4b}-\bigg( i+l+\frac{3}{2}\bigg)\right), \omega =l+1; \gamma = l+\frac{3}{2};\; \tilde{\varepsilon }= mr;\; z=\frac{b}{2}\mu r^2 \right) \nonumber\\
&&= \frac{\Gamma (\gamma +\beta _0)}{\Gamma (\gamma )} \Bigg\{\sum_{i_0=0}^{\beta _0 } \frac{(-\beta _0)_{i_0}}{(1)_{i_0}(\gamma)_{i_0}}z^{i_0}+  \Bigg\{\sum_{i_0=0}^{\beta _0 } \frac{(i_0+\frac{\omega }{2})}{(i_0+\frac{1}{2})(i_0-\frac{1}{2}+\gamma )} \frac{(-\beta  _0)_{i_0}}{(1)_{i_0}(\gamma)_{i_0}}\nonumber\\
&&\times  \sum_{i_1=i_0}^{\beta _1} \frac{(-\beta _1)_{i_1}(\frac{3}{2})_{i_0}(\gamma +\frac{1}{2})_{i_0}}{(-\beta _1)_{i_0}(\frac{3}{2})_{i_1}(\gamma +\frac{1}{2})_{i_1}} z^{i_1} \Bigg\} \tilde{\varepsilon }
+ \sum_{n=2}^{\infty } \Bigg\{ \sum_{i_0=0}^{\beta _0} \frac{(i_0+\frac{\omega }{2})}{(i_0+\frac{1}{2})(i_0-\frac{1}{2}+\gamma)} \frac{(-\beta _0)_{i_0}}{(1)_{i_0}(\gamma )_{i_0}} \nonumber\\
&&\times \prod _{k=1}^{n-1} \Bigg\{ \sum_{i_k=i_{k-1}}^{\beta _k} \frac{(i_k+\frac{\omega }{2}+\frac{k}{2})}{(i_k+\frac{1}{2}+\frac{k}{2})(i_k-\frac{1}{2}+\gamma + \frac{k}{2})} \frac{(-\beta _k)_{i_k}(1+\frac{k}{2})_{i_{k-1}}(\frac{k}{2}+\gamma )_{i_{k-1}}}{(-\beta _k)_{i_{k-1}}(1+\frac{k}{2})_{i_k}(\frac{k}{2}+\gamma )_{i_k}}\Bigg\} \nonumber\\
&&\times \sum_{i_n= i_{n-1}}^{\beta _n} \frac{(-\beta _n)_{i_n}(1+\frac{n}{2})_{i_{n-1}}(\frac{n}{2}+\gamma )_{i_{n-1}}}{(-\beta _n)_{i_{n-1}}(1+\frac{n}{2})_{i_n}(\frac{n}{2}+\gamma )_{i_n}} z^{i_n}\Bigg\} \tilde{\varepsilon }^n \Bigg\}
\label{eq:71}
\end{eqnarray}
Put (\ref{eq:71}) in $\Psi (r) = \exp\left(-\frac{b}{4}\left(r+\frac{2m}{b}\right)^2\right) r^l y(r)Y_l^{m^{\star}}(\theta ,\phi)$. The wave function for the spin free Hamiltonian involving only scalar potential for the $q-\bar{q}$ system  is given by
\begin{eqnarray}
\Psi(r) &=& N  r^l \exp\left(-\frac{b}{4}\left(r+\frac{2m}{b}\right)^2\right) QW_{\beta _i}\left( \beta _i= \frac{1}{2}\left( \frac{E^2}{4b}-\bigg( i+l+\frac{3}{2}\bigg)\right), \omega =l+1\right.\nonumber\\
&&;\left. \gamma = l+\frac{3}{2}; \tilde{\varepsilon }= mr;\; z=\frac{b}{2}\mu r^2 \right) Y_l^{m^{\star}}(\theta ,\phi) 
\label{eq:73}
\end{eqnarray}
N is normalized constant. Energy $E^2$ is 
\begin{equation}
E^2= 4b\left( 2\beta _i + i+l+\frac{3}{2} \right) \hspace{1cm} \mathrm{where}\; i,\beta _i \in \mathbb{N}_{0} \nonumber
\end{equation}
The GCH function with three recursive coefficients has infinite eigenvalues.

\section{Conclusion}
Any special functions with two recursive coefficients (such as Bessel, Legendre, Kummer, Laguerre, hypergeometric, Coulomb wave functions, etc) only have one eigenvalue for the polynomial case. However the GCH polynomial with three recursive coefficients has infinite eigenvalues that make $B_n$'s term terminated as we see (\ref{eq:60}), (\ref{eq:65}) and (\ref{eq:73}). 

I show power series expansions in closed forms of GCH equation in this paper. As we see analytic power series expansions of GCH equation by applying 3TRF \cite{Chou2012b}, denominators and numerators in all $B_n$ terms arise with Pochhammer symbol: the meaning of this is that the analytic solutions of GCH equation with three recursive coefficients can be described as hypergoemetric function in a strict mathematical way. Since this function is described as hypergeometric function, we can transform this function to other well-known special functions having two term recurrence relation: understanding the connection between other special functions is important in the mathematical and physical points of views as we all know.

In my next paper I derive the integral representation of GCH equation including all higher terms of $A_n$'s by applying 3TRF \cite{Chou2012b}. From integral forms of the GCH equation, we can investigate how these functions are associated with other well known special functions such as Bessel, Laguerre, Kummer, hypergeometric functions, etc. And I show generating functions for the GCH polynomial which makes $B_n$ term terminated. Generating functions are really useful in order to derive orthogonal relations, recursion relations and expectation values of any physical quantities as we all recognize; i.e. the normalized wave function of hydrogen-like atoms and expectation values of its physical quantities such as position and momentum.
\vspace{5mm}

\section{Series ``Special functions and three term recurrence formula (3TRF)''} 

This paper is 9th out of 10.
\vspace{3mm}

1. ``Approximative solution of the spin free Hamiltonian involving only scalar potential for the $q-\bar{q}$ system'' \cite{Chou2012a} - In order to solve the spin-free Hamiltonian with light quark masses we are led to develop a totally new kind of special function theory in mathematics that generalize all existing theories of confluent hypergeometric types. We call it the Grand Confluent Hypergeometric Function. Our new solution produces previously unknown extra hidden quantum numbers relevant for description of supersymmetry and for generating new mass formulas.

2. ``Generalization of the three-term recurrence formula and its applications'' \cite{Chou2012b} - Generalize three term recurrence formula in linear differential equation.  Obtain the exact solution of the three term recurrence for polynomials and infinite series.

3. ``The analytic solution for the power series expansion of Heun function'' \cite{Chou2012c} -  Apply three term recurrence formula to the power series expansion in closed forms of Heun function (infinite series and polynomials) including all higher terms of $A_n$'s.

4. ``Asymptotic behavior of Heun function and its integral formalism'', \cite{Chou2012d} - Apply three term recurrence formula, derive the integral formalism, and analyze the asymptotic behavior of Heun function (including all higher terms of $A_n$'s). 

5. ``The power series expansion of Mathieu function and its integral formalism'', \cite{Chou2012e} - Apply three term recurrence formula, analyze the power series expansion of Mathieu function and its integral forms.  

6. ``Lam\'{e} equation in the algebraic form'' \cite{Chou2012f} - Applying three term recurrence formula, analyze the power series expansion of Lam\'{e} function in the algebraic form and its integral forms.

7. ``Power series and integral forms of Lam\'{e} equation in Weierstrass's form and its asymptotic behaviors'' \cite{Chou2012g} - Applying three term recurrence formula, derive the power series expansion of Lam\'{e} function in   Weierstrass's form and its integral forms. 

8. ``The generating functions of Lam\'{e} equation in Weierstrass's form'' \cite{Chou2012h} - Derive the generating functions of Lam\'{e} function in   Weierstrass's form (including all higher terms of $A_n$'s).  Apply integral forms of Lam\'{e} functions in   Weierstrass's form.

9. ``Analytic solution for grand confluent hypergeometric function'' \cite{Chou2012i} - Apply three term recurrence formula, and formulate the exact analytic solution of grand confluent hypergeometric function (including all higher terms of $A_n$'s). Replacing $\mu $ and $\varepsilon \omega $ by 1 and $-q$, transforms the grand confluent hypergeometric function into Biconfluent Heun function.

10. ``The integral formalism and the generating function of grand confluent hypergeometric function'' \cite{Chou2012j} - Apply three term recurrence formula, and construct an integral formalism and a generating function of grand confluent hypergeometric function (including all higher terms of $A_n$'s). 
\section*{Acknowledgment}
I thank Bogdan Nicolescu. The discussions I had with him on number theory was of great joy.  
\vspace{3mm}

\bibliographystyle{model1a-num-names}
\bibliography{<your-bib-database>}

\end{document}